# Metastable polymorphic phases in monolayer TaTe$_2$

*Iolanda Di Bernardo\*‡ [1,2,3,4], Joan Ripoll-Sau,‡,[1,4] Fabian Calleja,[4] Cosme G. Ayani,[1,4] Rodolfo Miranda,[1,4,5,6] Jose Angel Silva-Guillén,[4] Enric Canadell,[7] Manuela Garnica,\*[4,5] Amadeo L. Vázquez de Parga[1,4,5,6]*

[1] Departamento de Física de la Materia Condensada, Universidad Autónoma de Madrid, 28049 Madrid, Spain

[2] ARC Centre of Excellence in Future Low-Energy Electronics Technologies, Monash University, 3800 Victoria, Australia

[3] School of Physics and Astronomy, Monash University, 3800 Victoria, Australia

[4] Instituto Madrileño de Estudios Avanzados en Nanociencia (IMDEA-Nanociencia), 28049 Madrid, Spain

[5] Instituto Nicolás Cabrera, Universidad Autónoma de Madrid, 28049 Madrid, Spain

[6] Condensed Matter Physics Center (IFIMAC), Universidad Autónoma de Madrid, 28049 Madrid, Spain

[7] Institut de Ciència de Materials de Barcelona, ICMAB-CSIC, Campus UAB, 08193 Bellaterra, Spain

Polymorphic phases and collective phenomena – such as charge density waves (CDWs) - in transition metal dichalcogenides (TMDs) dictate the physical and electronic properties of the material. Most TMDs naturally occur in a single given phase, but the fine-tuning of growth conditions via methods like molecular beam epitaxy (MBE) allows to unlock otherwise inaccessible polymorphic structures. Exploring and understanding the morphological and electronic properties of new phases of TMDs is an essential step to enable their exploitation in technological applications. Here, we use scanning tunnelling microscopy to map MBE-grown monolayer TaTe$_2$. We report the first observation of the 1H polymorphic phase, coexisting with the 1T, and demonstrate that their relative coverage can be controlled by adjusting synthesis parameters. Several super-periodic structures, compatible with CDWs, are observed to coexist on the 1T phase. Finally, we provide theoretical insight on the delicate balance between Te...Te and Ta-Ta interactions that dictates the stability of the different phases. Our findings demonstrate that TaTe$_2$ is an ideal platform to investigate competing interactions, and indicate that accurate tuning of growth conditions is key to accessing metastable states in TMDs.

## 1. Introduction

Transition metal dichalcogenides (TMDs) have been intensively studied for the variety of physical properties they offer.[1,2] They are layered compounds with the formula MX$_2$, where M is a transition metal and X a chalcogen, characterized by weak, mostly van der Waals-like interactions between planes. According to the chalcogen coordination geometry within the unit cell, ultimately related to the *d*-electron count on the metal, TMDs are classified in different phases or polymorphs (H, T, T', distorted T', etc.).[1,2] Different stacking orders of TMD layers give rise to different polytypes (1T, 2H, 3R, etc.). The trigonal prismatic phase is known as 1H (in monolayers) and has D$_{3h}$ point symmetry. It is a hexagonal phase, with the chalcogens vertically aligned in the unit cell (**Figure 1a**). In the 1T octahedral phase, characterized by D$_{3d}$ symmetry, one of the chalcogen planes glides with respect to the other (**Figure 1b**). It is worth noting that the top layer of chalcogens in a 1T polymorph still exhibits hexagonal symmetry.

The H and T phases are the most stable among group VI and group IV TMDs respectively.[3] In group V TMDs, both H and T phases are found and it is even possible to induce the transition between the two phases.[4,5] Reducing the thickness down to a monolayer (ML) and tailoring the growth conditions (temperature, strain, substrate geometry) can result in TMD films with a polymorphic phase - and physical properties - strikingly different from the bulk one.[6,7] For example, an indirect to direct bandgap transition has been observed by thinning down 2H-MoS$_2$[8] and 2H-MoSe$_2$[9] to the monolayer; the emergence of an excitonic insulator state has been reported in monolayer 1T'-WTe$_2$;[10,11] IrTe$_2$, metallic in the bilayer, dimerizes in a distorted ($2x1$) 1T structure as a ML and undergoes a metal-to-insulator transition;[6] Ising superconductivity is observed in ML NbSe$_2$.[12]

Group V TMDs (M= V, Nb, Ta) have played a significant role in the unraveling of CDW phenomena. Most of the initial studies focused on CDWs in group V sulfides and selenides, with tellurides arousing real interest only in the last couple of years. The very diffuse nature of the tellurium orbitals may cause a considerably large electron transfer from the chalcogens to the transition metal atoms,[13] leading to remarkable differences with their sulfides and selenides counterparts. The absence of inter-layer Te…Te interactions in ML can result in a substantial variation of the electronic transfer compared to a multi-layer system. This drove the interest to prepare and characterize ML group V transition metal ditellurides.

Bulk TaTe$_2$ is metallic[14] and exhibits a distorted 1T' structure at room temperature,[14,15] with surface Ta atoms arranged in double Ta zigzag trimer chains (see **Figure S1a**). This peculiar structure is formed as a result of the abovementioned large Te to Ta electron transfer.[16,17] Because of the trimeric zigzag chains, each layer of 1T'-TaTe$_2$ can be thought of as a $(3x1)$ reconstruction of an ideal, undistorted 1T structure. Bulk 1T'-TaTe$_2$ undergoes a structural deformation around $170 K$,[18] below which the zigzag trimer chains rearrange and the periodicity along the chains becomes three times larger. Consequently, below $170 K$ the layers can be described as a $(3x3)$ reconstruction of 1T-TaTe$_2$ (see **Figure S1b**). The formation of a CDW at low temperatures and the emergence of a $(\sqrt{19}x\sqrt{19})$ periodicity (**Figure S1c**) has also been known for quite some time.[19,20] The metallic character is kept across the transition, although anomalies in the resistivity, magnetic susceptibility and specific heat capacity are detected.[18]

The first molecular beam epitaxy (MBE) growth of thin films of TaTe$_2$ on bilayer graphene/SiC was recently reported.[21] The authors observed a 1T phase (rather than the distorted 1T') for thicknesses up to 8MLs, and irreversible CDW transitions in the ML as a function of the substrate annealing temperature. For high substrate temperatures and thicknesses above the ML, the most stable super-structure was found to be the $(\sqrt{19}x\sqrt{19})$ CDW. A photoemission-based work also recently reported the formation of a 2H reconstruction on the topmost layer of a distorted 1T'-TaTe$_2$ bulk crystal.[22] These studies suggest the possibility of phase coexistence and phase tunability with thickness, as previously reported for other TMDs.[7,11,23–25]

In this work, we report the growth of sub-monolayer TaTe$_2$ on a graphene/Ir(111) (Gr/Ir(111)) substrate via MBE. We observe the coexistence of the two metastable 1H (trigonal prismatic) and 1T (octahedral) phases and investigate their properties via scanning tunnelling microscopy (STM) and spectroscopy (STS) at room temperature (RT) and at $77 K$. At both temperatures, the octahedral phase exhibits periodic super-structures, compatible with the ones observed in bulk crystals. The 1H phase, never reported before, undergoes a restructuring of its super periodicity between RT and $77 K$. We also demonstrate tuning of the relative phase coverage by controlling the growth conditions and extract information about the Te desorption energy from the surface by a simple adsorption/desorption model. The relative stability of the different crystallographic phases and periodicities found in the experiments are rationalized by means of first-principles density functional theory (DFT) calculations. The combination of MBE growth, in-situ STM characterization and first-principles calculations is a powerful strategy to obtain and investigate metastable polymorphic phases with predicted new properties not present in the naturally occurring crystals.[14,26,27]

## 2. Results

We investigate the structure of MBE-grown $TaTe_2$ on Gr/Ir(111) in the sub-monolayer regime. **Figure 1c** shows a large scale ($300\ nm\ x 300\ nm$) STM topographic image of $TaTe_2$ islands (bright) on Gr/Ir(111) (dark). The islands appear flat, slightly dendritic and cover about 60% of the surface. A line profile across an island step edge (blue line in Figure 1c) reveals a step height of 6.9 Å, compatible with the thickness of a single TMD layer.[23,28] We can tune the $TaTe_2$ coverage by controlling growth time and elemental precursors ratio, and we observe minimal bilayer formation even at coverages as high as 80%. **Figure 1d** shows a close-up ($15\ nm\ x\ 15\ nm$) on the surface. Visible in the bottom part of the image is the well-known moiré pattern of graphene on Ir(111).[25] In the top part of the image, we observe a $TaTe_2$ island with two different reconstructions separated by a line defect. On both sides of the island the atomic arrangement presents a hexagonal symmetry, with the left-hand side additionally showing a super periodic pattern. Neither side, however, shows a reconstruction compatible with the double zigzag "ribbon chain" corresponding to the distorted 1T' phase observed on the surface of bulk $TaTe_2$.[17,29,30]

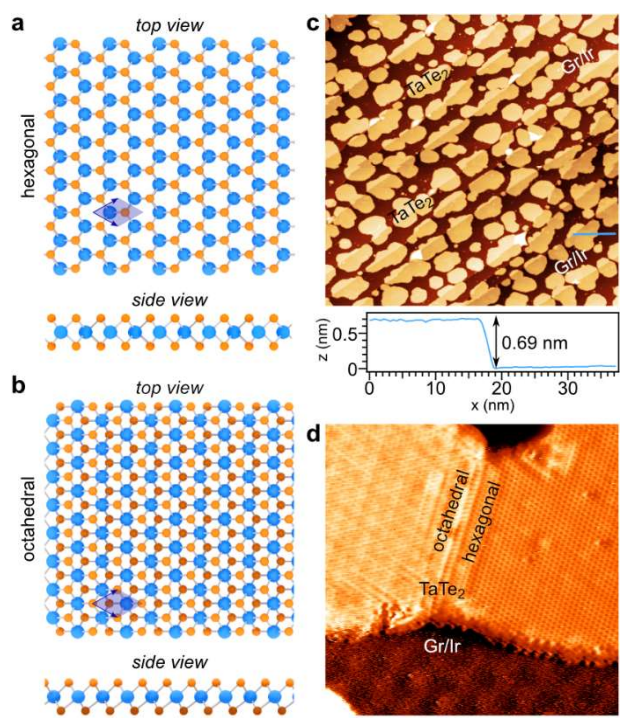

**Figure 1:** Ball-stick model of the hexagonal **(a)** and octahedral **(b)** polymorphs, top and side view. Blue and orange balls represent the Ta and Te atoms, respectively; in (b) the different shade of orange accounts for the two inequivalent chalcogen sites. Blue arrows enclose the shaded area corresponding to one unit cell. **(c)** Large scale ($300\ nm\ x\ 300\ nm$) topographic image showing about 60% TaTe$_2$ ML coverage (bright islands) on Gr/Ir(111) (dark substrate). A line profile across an island step edge (blue dashed line in the top panel) reveals a height of 6.9 Å. Scalebar: $50\ nm$; $I = 0.1\ nA$; $V = 1V$. **(d)** Zoom-in ($15\ nm\ x\ 15\ nm$) topographic image of a TaTe$_2$ island showing two different reconstructions. To enhance features, the $z$ signal is mixed with its derivative. $I = 0.3\ nA$; $V = 0.4\ V$. All data acquired at room temperature.

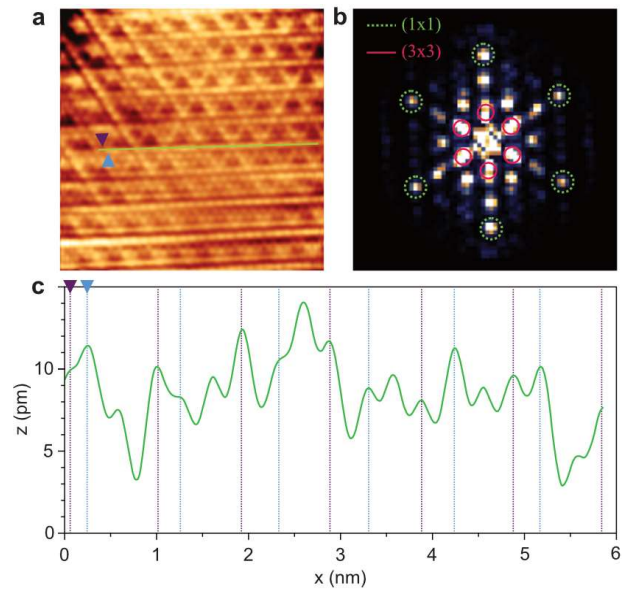

**Figure 2:** (a) ($8\ nm\ x\ 8\ nm$) image of an octahedral area ($I = 2nA$; $V = -0.5V$). **(b)** 2D fast Fourier transform of the image shown in panel (a). **(c)** Line profile acquired along the green dashed line in (a). The purple and blue triangles in panels (a) and (c) indicate the starting positions of the consecutive maxima in the line profile. In (c), light blue and purple dashed lines serve as a guide to the eye to track the position of one on every three maxima, highlighting the incommensurability of the system. In (b), the spots corresponding to the (1x1) and quasi-(3x3) periodicities are highlighted in dashed green and solid red, respectively. All measurements were carried out at room temperature.

**Figure 2** shows a high-resolution STM topographic image obtained on an island showing the octahedral phase. The pattern in Figure 2a closely resembles the one reported for the topmost layer of bulk 1T'-TaTe$_2$ at $77K$,[29] and was never observed before at room temperature. A threefold-symmetric pattern is seen in the top part of the image, while the reconstruction on the lower part exhibits a preferential symmetry axis. This pattern was ascribed to an incommensurate CDW, resulting from the competition between the hexagonal $(3x3)$ and the rectangular $(3\sqrt{3}/2\ x4)R30°$ sharing a symmetry axis along the direction of the stripes.[29] The corresponding fast Fourier transform (FFT) pattern (Figure 2b) exhibits two distinct sets of spots corresponding to the (1x1) periodicity (dashed green circles) and to an incommensurate quasi-(3x3) (solid red circles). To verify the incommensurability of the pattern, we take a line profile across the image (green line in Figure 2a) and track the position of one in every three maxima starting from two consecutive maxima (highlighted in blue and purple, respectively) along this line. The profile is reported in Figure 2c, with the position of the maxima marked with light blue and purple dashed vertical lines. The distance between the maxima varies along the profile, confirming the incommensurability of the system. The room temperature evolution of this pattern as a function of the tip-sample bias is shown in **Figure S2**. At RT we also observe small patches of a flower-like reconstruction, similar to the one recently reported by Hwang and coworkers,[21] and a $(3x1)$ reconstruction (see **Figure S3**), both discussed below. The existence of CDWs has been reported for other ditellurides ($VTe_2$,[31] $TiTe_2$,[32] $NbTe_2$[33]), and demonstrated for 1T-TaTe$_2$ film thickness up to 8 ML.[21] We, therefore, tentatively attribute these super periodicities to the presence of CDWs at room temperature. The average step height of the octahedral islands is $732 \pm 4$ pm, obtained by averaging over forty profiles acquired with different bias voltages across different sample preparations (see **section S3** for details).

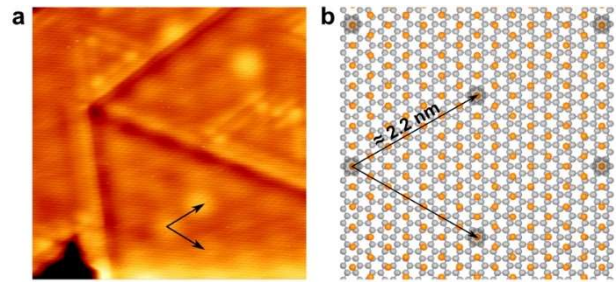

**Figure 3:** (a) ($15\ nm\ x\ 15\ nm$) room temperature image of a 1H area ($I = 3nA; V = 1V$). **(b)** Cartoon model of the moiré lattice of TaTe$_2$ on graphene; grey and orange balls represent carbon and topmost chalcogen layer atoms, respectively.

Coexisting with the octahedral phase we observe another hexagonal phase shown in **Figure 3a**. The average step height of these islands is $681 \pm 6\ pm$, consistently smaller than the one measured on the islands with the octahedral phase (see section S3 for details). The smaller step height of the islands can be understood if it is assumed that crystallized islands in different polymorphs co-exist on the surface. In particular, the H-phase step edges are expected to be slightly lower than the octahedral ones because the partial distortion of the latter induces a buckling of the chalcogen atoms. These islands also exhibit a superlattice modulation, but its periodicity and pattern do not change as a function of tip-sample bias (**Figure S4**). We attribute this intensity modulation to the moiré pattern between the graphene substrate and an H-phase island. A cartoon model of the moiré is reported in **Figure 3b**. Assuming a graphene lattice constant of 2.4 Å and a 1H-TaTe$_2$ constant of 3.6 Å,[26] the (incommensurate) moiré lattice constant is expected to be ~2.2 *nm*, matching experimental observations. Corroborating the attribution to an H phase is the observation of mirror twin boundaries (MTB) in this type of islands. MTBs have primarily been reported for other H-TMDs grown on hexagonal substrates,[34–37] and at the interface between octahedral and H islands.[38] MTBs form when the adlayer has a threefold symmetry (120°) while the substrate has a sixfold (60°) one: two TMD 60°-rotated domains are equivalent to two mirrored ones, and their junction will form a MTB.[37,39] The relative coverage of the H and octahedral areas ratio on the sample can be tuned as a function of substrate temperature and growth time (see discussion section).

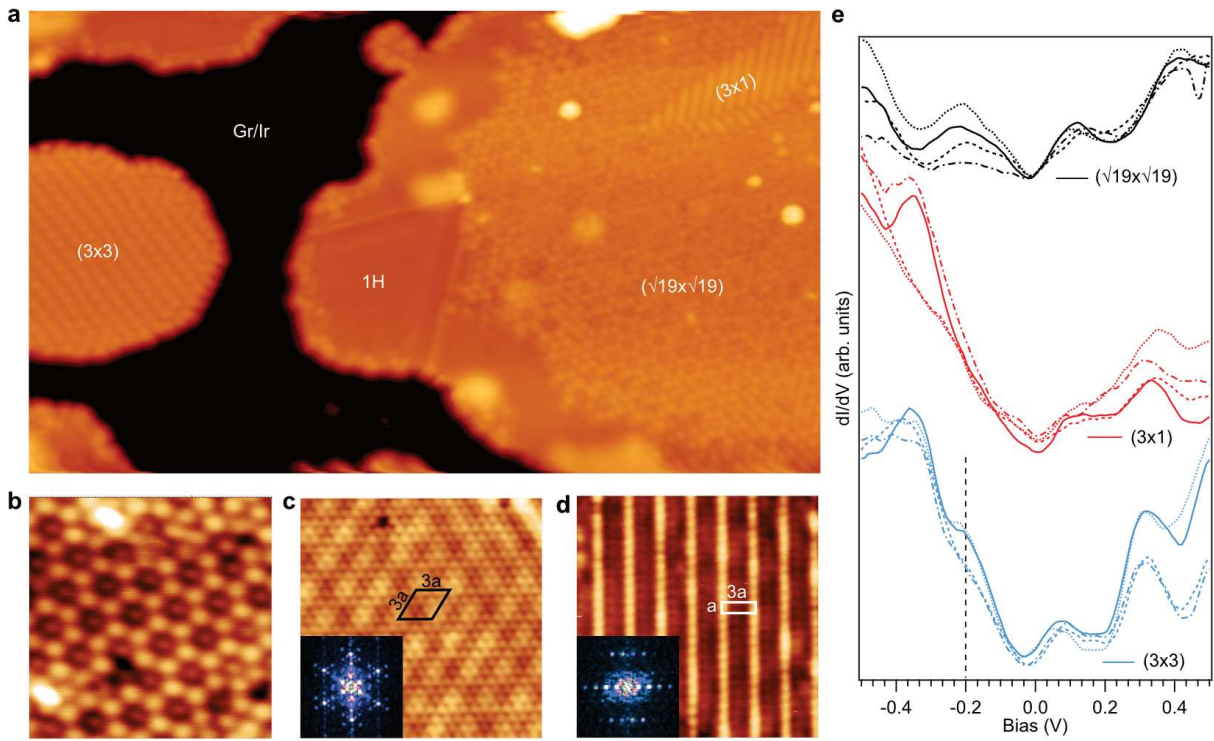

**Figure 4: (a)** $(60\ nm\ x\ 35\ nm)$ topography showing the coexistence of two distinct phases and multiple super periodic patterns ($I = 0.3\ nA; V = -1V$). Close-up micrographs of areas exhibiting reconstructions compatible with: **(b)** $(\sqrt{19}x\sqrt{19})$, **(c)** $(3x3)$, **(d)** $(3x1)$ superlattices. (b) $(8\ nm\ x\ 8\ nm)$, $I = 0.1\ nA, V = -1V$. (c) $(7\ nm\ x\ 7\ nm)$, $I = 1nA; V = 10\ mV$. (d) $(7\ nm\ x\ 7\ nm), I = 1nA; V = -0.1V$. **(e)** STS spectra acquired on the different periodic reconstructions. Spectra of the same color but different line styles are acquired on the same area but different spots. All data acquired at $77K$.

**Figure 4** reports the different polymorphs and super periodicities observed at $77K$ on the sample prepared under similar conditions than before. The flatter-looking areas, generally enclosed by MTBs, are attributed to the hexagonal phase, and described below. The rest is assigned to the octahedral areas, which reveal different superstructures, as seen in panels (b)-(d). The flower-like structure in Figure 4b, also observed at RT, has recently been attributed to the $(\sqrt{19}x\sqrt{19})$CDW[21] originally reported in bulk TaTe$_2$.[19,20] An alternative explanation identifies this superstructure as a chalcogen-deficient stoichiometric phase, like the one reported for Mo$_5$Te$_8$.[40] We find this pattern to be predominant for higher coverages and higher growth temperatures. This would suggest that it is the most thermodynamically stable among the superlattices for the growth parameters explored in this study. Further investigations, based on stoichiometrically accurate measurements (such as XPS or EDX, and beyond the scope of this work), are necessary to univocally establish the correct interpretation.

On other octahedral-phase islands (left of Figure 4a), we distinguish either a $(3x3)$ reconstruction (Figure 4c), appearing like a hexagonal close packed lattice of bright bumps on the large-scale images, or a $(3x1)$ reconstruction (Figure 4d). The latter is usually embedded within larger areas of $(\sqrt{19}x\sqrt{19})$, and typically exhibits a fish-scale pattern, with adjacent domains rotated by 120°. Notably, a competition between the triple-axis $(3x3)$ and the single axis $(3x1)$ periodicities was reported to be at the origin of the "irregular" CDW observed on the surface of bulk TaTe$_2$ at $77K$[29] and theoretical calculations predict the existence of both $(3x1)$ and $(3x3)$ CDWs in ML 1T-TaTe$_2$.[26] While these observations would support the attribution of the $(3x1)$ and $(3x3)$ patterns to CDWs, based on our STM measurements it is difficult to disentangle whether the super-periodicities are to be univocally ascribed to charge modulation or an actual lattice reconstruction.

We report the STS spectra on the three differently patterned areas of panels (b)-(d) at $77K$ in Figure 4e. We observe a series of common features, like the presence of a dip in the density of states in proximity of the Fermi level – which does not, in any case, go to zero and open a bandgap (see section S4 for a detailed discussion). These observations point to the coexistence of multiple CDWs on the surface at the probed temperatures, rather than the conversion of one into another at given growth and annealing conditions.

The introduction of a new periodicity in the system below a certain temperature should reflect on the band structure of the system. The phase transition at $170K$ in bulk 1T'-TaTe$_2$ has recently been studied via ARPES: at low temperatures, a suppression of the spectral weight, associated with the formation of minigaps, has been observed at binding energies of $-100$ and $-270\ meV$. Such suppression is attributed to the band folding effect induced by the formation of the $(3x3)$ periodicity below $170K$. As a consequence of these dips, the DOS appears to have a peak at $\sim -200\ meV$. The blue spectra in Figure 4e, showing STS measured on areas of our sample with $(3x3)$ periodicity, show indeed a shoulder at about $-200meV$ (black vertical dashed line in Figure 4e). STS spectra measured on the areas with a $(3x1)$ periodicity, on the other hand, do not show such feature, in good agreement with the ARPES data. The other spectral characteristics, common to all the 1T areas of the surface, can be ascribed to the shape of the band structure at the Fermi level in the T phase of TaTe$_2$.[27]

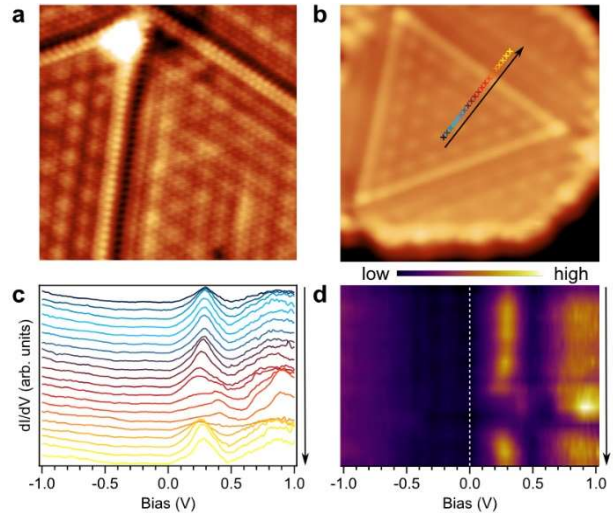

**Figure 5**: 1H phase characterization at $77K$. **(a)** Atomically resolved $(10\ nm\ x\ 10\ nm)$ image of a 1H area showing MTBs and the $(3x3)$ CDW. $I = 200pA;\ V = 10mV$. **(b)** $(6\ nm\ x\ 6\ nm)$ image showing an MTB loop. $I = 100pA;\ V = -1V$. **(c)** STS spectra acquired along the line in panel (b); the spectra are color-coded and vertically offset for clarity. **(d)** Heat-map corresponding to the spectra in panel (c).

At $77K$ we observe a $(3x3)$ modulation on the hexagonal phase as well, as seen in **Figure 5a** and **b**. The enclosure within MTBs with a 60° angle between them supports the attribution to an H phase rather than a T one. Closed MTB triangular loops, like the one in Figure 5b, have been reported to originate from slightly chalcogen-deficient growth conditions.[37] The spectral weight distribution in STS data acquired on the 1H-TaTe$_2$ phase showing the $(3x3)$ reconstruction (**Figure 5c**) is very different from that acquired on the octahedral phase (Figure 4e). The spectra reflect the metallic character of the 1H phase, with a pronounced feature at positive bias (empty states). This feature, located at $\approx 270\ meV$, could originate from a band edge located at the Γ point according to the calculated band structure,[27] like in NbSe$_2$.[41,42] We perform line-profile STS in the direction perpendicular to the MTB, presented as a stack in Figure 5c and as a heat map in **Figure 5d**. The black arrow in panels (b)-(d) points in the direction of the scan acquisition. The metallicity of the sample is preserved across the MTB, with a charge redistribution in the empty states seen in correspondence of the boundary itself.

## 3. Discussion

The possibility to grow TaTe$_2$ MLs in both trigonal prismatic and octahedral coordination, as well as the coexistence of different CDW patterns, are striking and call for a rationalization. We thus carried out first-principles DFT calculations on different phases of TaTe$_2$ to estimate their stabilities (see **Tables 1** and **2**).

The reliability of the calculations was tested by first looking at TaTe$_2$ bulk. The ground state of bulk TaTe$_2$ is found to be the $(3x3)$ phase, with the 1T' (i.e. $(3x1)$) only $6\ meV/f.u.$ higher in energy. By contrast, the two hexagonal 1T and 2H phases are found at considerably higher energies (160 and $128\ meV f.u.^{-1}$, respectively) (see Table 1). These results are in excellent agreement with the experimental observation that bulk TaTe$_2$ crystals exists only in the distorted 1T' phase at room temperature and distorts to the (3x3) one at low temperature, lending credit to the calculations for the ML.

We considered self-standing monolayers of the 1T'-$(3x3)$ and 1T'-$(3x1)$ phases as well as other possible superstructures of the 1T and 1H phases: $(\sqrt{7}x\sqrt{7})$, $(\sqrt{13}x\sqrt{13})$ and $(\sqrt{19}x\sqrt{19})$ for 1T; $(3x3)$ for 1H. Overall, the more stable phase is found to be the 1T'-$(3x3)$, followed by the 1T'-$(3x1)$ at a relatively small energy difference of $12\ meV/f.u.$. The other super-periodicities are all found at higher energies (see Table 2). The 1T and 1H phases remain considerably higher in energy in the monolayer regime (see Table 2).

**Table 1:** Relative energies, $\Delta E, (meV\ f.u.^{-1})$ for different TaTe$_2$ phases in bulk.

|  | Octahedral | | | Hexagonal |
|---|---|---|---|---|
| **Phase** | $3x3$ | $3x1$ | 1T | 2H |
| ΔE | 0 | +6 | +160 | +118 |

Table 2: Relative energies, $\Delta E, (meV\ f.u.^{-1})$ for different ML-TaTe$_2$ phases.

| | Octahedral | | | | | | Hexagonal | |
|---|---|---|---|---|---|---|---|---|
| **Phase** | 3x3 | 3x1 | √7x√7 | √13x√13 | √19x√19 | 1T | 1H | 3x3 |
| ΔE | 0 | +12 | +82 | +17 | +50 | +131 | +122 | +100 |

Table 3: Relative energies between the 1H (hexagonal) and 1T (octahedral) phases, $\Delta E_{(1H-1T)}, (meV\ f.u.^{-1})$ for different MTe$_2$ compounds.

| Element (M) | V | Nb | Ta |
|---|---|---|---|
| ΔE$_{1H-1T}$ | +100 | +6 | -9 |

The theoretical results provide some useful hints concerning the stability of different ML TaTe$_2$ phases:

(i) The large stabilization of 1T vs 1H when going from bulk to ML shows that inter-layer interactions have an important differential effect in stabilizing certain phases of TaTe$_2$.

(ii) While at RT bulk TaS$_2$ and TaSe$_2$ are found in 1T/2H forms, TaTe$_2$ is found in the 1T' octahedral structure. Our calculations suggest that the strong preference for the octahedral coordination is due to a strong driving force towards a 1T'-type clustering, leading to short Ta-Ta bonds in both bulk and ML. Ta-Ta clustering in H type layers provides a stabilization which is only 5-6 times weaker.

(iii) Comparing the results in Tables 1 and 2 we observe that, when the inter-layer Te…Te interactions are quenched (in the ML), the stabilization of the 1T- (3x1) phase with respect to the 1T is weaker. This supports previous suggestions that the Te to Ta electron transfer, which is maximized in the bulk, is the major factor behind the distortion toward the 1T' structure.[16,17]

(iv) The suppression of the inter-layer Te…Te interactions in the ML increases the stability of the (3x3) phase with respect to the 1T' (3x1). This means that the *inter*-layer Te…Te interactions are not the only factor to consider in understanding the relative stability of these phases: the inter-layer Te…Te interactions favor the distortion from 1T towards the 1T' (3x1) phase, while the intra-layer Te…Te interactions favor the subsequent distortion toward the (3$x$3) phase. In other words, the decompression effect of removing adjacent layers is also responsible for changes in the intra-layer Te…Te interactions.

(v) While the 1T phase is preferred over the 1H in ML VTe$_2$ and ML NbTe$_2$, such preference is reversed for ML TaTe$_2$ (see **Table 3**). In any way, this means that the metal-to-metal interactions also influence in a non-negligible way the stability of the TaTe$_2$ phases.

(vi) A recent MBE study[21] supports our conclusion that a phase compatible with a ($\sqrt{19}x\sqrt{19}$) CDW could be prepared under certain conditions. As shown in Table 2, all ($\sqrt{m}x\sqrt{m}$) superstructures considered here have energy stabilities in-between those of the (3$x$3) and 1H phases observed in the present work. This suggests that ML phases like the ($\sqrt{7}x\sqrt{7}$) and ($\sqrt{13}x\sqrt{13}$) may also be prepared via MBE, by appropriately tuning the working conditions.

Taken together, these observations suggest that there is a delicate trade between Te…Te and Ta…Ta interactions, which govern the stability of different TaTe$_2$ ML phases. MBE may provide a very useful approach to prepare TaTe$_2$ phases differing by as much as ~100 $meV f.u.^{-1}$ in stability using appropriate growth conditions.

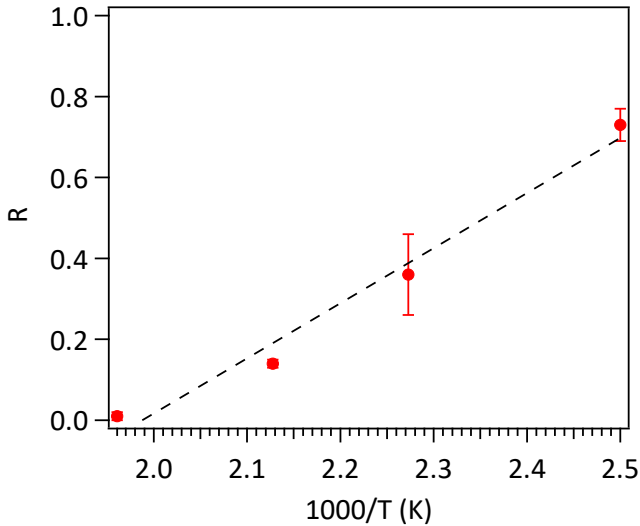

**Figure 6:** Phase tunability with temperature. Hexagonal:octahedral phase area ratio (R) as a function of the inverse of growth temperature. Data extracted from STM measurements acquired at room temperature.

We now discuss phase tunability. Usually, phase engineering of TMDs is achieved by changing the occupation of the *d* orbitals of the transition metal[43] (for example by alkali/salt intercalation) or by applying strain. Tuning of the growth temperature (therefore, of the adatom mobility on the substrate) and of the metal-chalcogen ratio have also been reported as efficient techniques to control the morphology and grain size in transition metal diselenides[28] and ditellurides.[25] We can phase-engineer ML TaTe$_2$ by tuning the growth temperature. Considering all super periodicities, we observe a larger octahedral:hexagonal area ratio for higher growth temperatures (**Figure 6** and **S6**) and longer growth times, in line with theoretical prediction that the octahedral phase is the most stable for the ML. Note that our experiments are apparently contradicting the results of Hwang et al.,[21] who observe only the 1T phase in their work. The authors used a similarly inert substrate (bilayer graphene on SiC, compared to our graphene on iridium) and reported a Ta:Te ratio during growth between 1:20 and 1:30, well within our 1:6 – 1:40 range. The lowest growth temperature reported in their work[21] however, is $535\ K$, larger than the highest growth temperature reported in this work ($510\ K$). Extrapolating from the plot in Figure 6, at temperatures higher than $510\ K$ the only phase expected is the 1T, coherently with the recent report.

The temperature-dependent 1H → 1T phase transition can be used to extract information about the tellurium desorption energy from the substrate surface, following the approach of Chen and coworkers.[44] We define the H:T ratio as $R = \frac{A_H}{A_T + A_H}$, calculated from the relative phase coverage for different growth temperatures estimated via STM image processing. We then plot R against the inverse of growth temperature and observe that the formation of the 1H phase (high R values) is favored at lower growth temperatures. The 1H phase is therefore promoted when tellurium adatoms linger on the surface for longer before being incorporated in an island or being desorbed. From the plot in Figure 5, R follows an Arrhenius model: $R \propto e^{E/k_B T} \approx E/k_B T$ where $E$ is the activation energy for the reaction. This activation energy can be related to the rate of adsorption/desorption of Te form the surface. In first order adsorption-desorption kinetics, neglecting the incorporation in islands, the rate of change of the adsorbate coverage $\vartheta$, proportional to R, is $\frac{d\vartheta}{dt} = F k_\alpha - \vartheta k_d$ where $F$ is the flux and the $k_{\alpha/d}$ are the adsorption/desorption coefficients. At steady state $\frac{d\vartheta}{dt} = 0$, so $\vartheta = F \frac{k_\alpha}{k_d}$. As $R \propto \vartheta$, by doing a linear fit on the plot we can extract an activation energy of 0.12 eV, value linked to the energy barrier necessary to desorb Te from the surface. This value is in good agreement with the 0.17 eV calculated to be necessary for desorption of Te from HOPG.[44] This plot demonstrates the important role of the growth temperature for tuning the phase of TMDs ML.

## 4. Conclusion

We have discovered the possibility of growing, via MBE, monolayers of the hexagonal and octahedral polymorphs of the TMD TaTe$_2$. These two metastable phases are not observed in bulk crystals. STM measurements reveal that the octahedral phase exhibits multiple coexisting – rather than mutually exclusive – CDW patterns at room temperature and $77 K$. STM and STS measurements also highlight the existence of a metallic 1H phase, never reported before, which supports the formation of CDW at cryogenic temperatures. Theoretical results demonstrate that a subtle trade between Te…Te (both inter- and intra-layer) and Ta…Ta interactions governs the stability of the different TaTe$_2$ ML phases. Tuning the relative coverage of one phase over the other is simply obtained by tuning of the growth temperature. Our combined experimental-theoretical study demonstrates that MBE is a simple but reliable method to obtain bulk-unstable phases and lateral homojunctions, as phases differing in stability by as much as ~100 $meV f.u.^{-1}$ can be prepared using this technique. Our synthesis method enables to access and study a very rich landscape of different phases and lattice super-modulations, never reported before for any other transition metal dichalcogenide.

## 5. Experimental Section/Methods

*MBE growth:* The gr/Ir(111) surface was prepared by 3 cycles of Ar+ ion sputtering at 1 $keV$ (pressure $3.6x10^{-6}$ $Torr$) followed by a flash annealing at 1600$K$. After last cycle, the surface was exposed to 33$L$ of ethylene while keeping the sample at 1440$K$. This well-established procedure[25] results is an atomically flat Ir(111) surface almost completely covered by a single layer of graphene. TaTe$_2$ growth was carried out on the gr/Ir(111) substrate by co-evaporating elemental Ta and Te from an e-beam (applied voltage = 1$kV$, flux= 12/24 $nA$) and a Knudsen cell (T= 600$K$), respectively. The flux ratio between elemental components was estimated by deposition of each of them individually on the substrate. The growth temperature ranged between 400 and 510K; at a substrate temperature of 440$K$ the film growth rate is estimated to be about 0.6 $ML\ h^{-1}$ by analysing substrate coverge via STM.

*STM-STS Measurements.* The morphology and electronic structure of the samples were measured using either a room temperature (RT) STM system, directly connected to the preparation chamber, or a low temperature (LT) system, where the STM head is in thermal equilibrium with a liquid nitrogen bath (base temperature =77$K$). For STS measurements a lock-in amplifier was used, where the modulation frequency was 721 $Hz$.

*First-principles calculations.* Numerical calculations were carried out using density functional theory (DFT)[45,46] as implemented in the SIESTA code.[47–49] We used the generalized gradient approximation (GGA), specifically, the revised Perdew-Burke-Ernzerhof functional for solids and surfaces PBEsol.[50] To describe the core electrons we used norm-conserving scalar relativistic pseudopotentials with non-linear core corrections.[51,52] The valence electrons were described with a split-valence double-ζ basis set including polarization functions.[53] The energy cutoff of the real space integration mesh was set to 500 Ry. To build the charge density (and, from this, obtain the DFT total energy and atomic forces), the reciprocal space was sampled with the Monkhorst-Pack scheme.[54] The number of k-points was optimized for two structures for both the bulk and single-layer systems, the 1T and 1T', resulting in grids of ($39 \times 39 \times 7$) and ($10 \times 42 \times 14$) for bulk and ($39 \times 39 \times 1$) and ($10 \times 42 \times 1$) for the single-layer case. For the other phases the grid was scaled accordingly. Both the lattice parameters as the atomic positions were optimized. The forces acting on the atoms were smaller than $0.04$ eVAng$^{-1}$.

**Supporting Information**

Supporting Information is available from the Wiley Online Library or from the author.

**Acknowledgements**

I. Di Bernardo and J. Ripoll-Sau contributed equally to this work. This work was supported by the Spanish Ministry of Science and Innovation (Grant no. PID2021-123776NB-C21, PGC2018-096955-B-C44 and PID2021-128011NB-I00) and the Comunidad de Madrid (Project S2018/NMT-4511, NMAT2D). IMDEA Nanociencia and IFIMAC acknowledge financial support from the Spanish Ministry of Science and Innovation 'Severo Ochoa' (Grant CEX2020-001039-S) and 'María de Maeztu' (Grant CEX2018-000805-M) Programme for Centres of Excellence in R&D, respectively. Financial support through the (MAD2D-CM)-MRR MATERIALES AVANZADOS-IMDEA-NC and (MAD2D-CM) MRR MATERALES AVANZADOS-UAM is acknowledged. M.G. has received financial support through the "Ramón y Cajal" Fellowship program (RYC2020-029317-I). I.D.B. acknowledges support from the Maria de Zambrano fellowship program and the MSCA Program (101063547-GAP-101063547), as well as from the FLEET Centre of Excellence, ARC Grant No. CE170100039. E.C. acknowledges support of the Spanish MICIU through the Severo Ochoa FUNFUTURE (CEX2019-000917-S) Excellence Centre distinction and Generalitat de Catalunya (2017SGR1506). J.A.S-G. thankfully acknowledges the computer resources at Marigold and the technical support provided by SCAYLE (RES-FI-2020-2-0040) as well as the computer resources of CCCUAM.

We demonstrated the formation of large scale TaTe$_2$ areas by molecular beam epitaxy with different polymorphs, some of them not reported before as the 1H phase. Our first-principles calculations rationalized these results and show that the different phases are a consequence of a delicate trade between Te…Te and Ta-Ta interactions.

**Metastable polymorphic phases in monolayer TaTe$_2$**

ToC figure

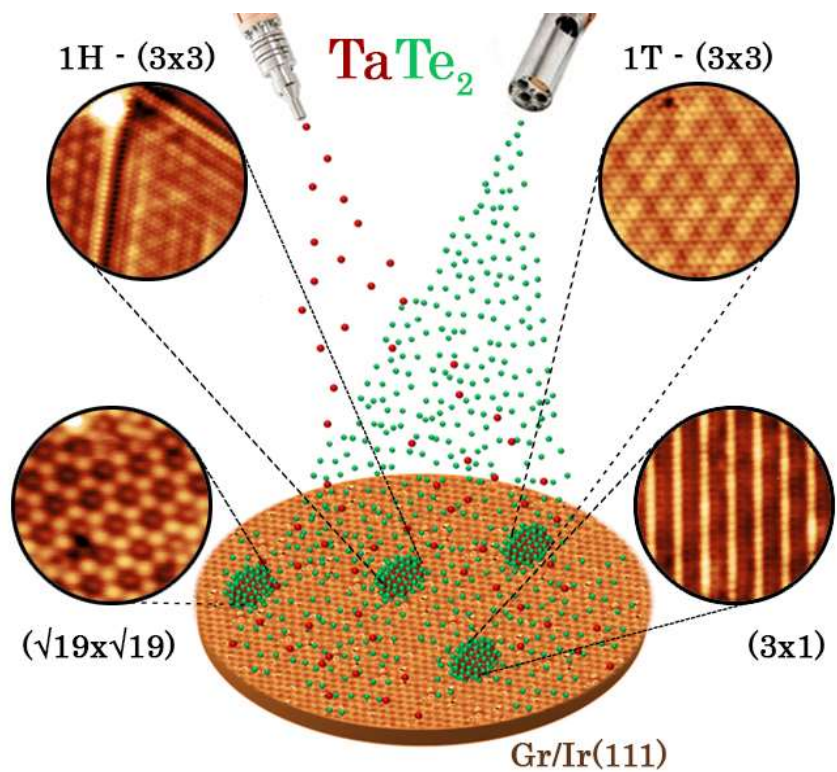

Supporting Information

**Metastable polymorphic phases in monolayer TaTe$_2$**

*Iolanda Di Bernardo\*‡ [1,2,3,4], Joan Ripoll-Sau,‡,[1,4] Fabian Calleja,[4] Cosme G. Ayani,[1,4] Rodolfo Miranda,[1,4,5,6] Jose Angel Silva-Guillén,[4] Enric Canadell,[7] Manuela Garnica,\*[4,5] Amadeo L. Vázquez de Parga[1,4,5,6]*

[1]Departamento de Física de la Materia Condensada, Universidad Autónoma de Madrid, 28049 Madrid, Spain

[2]ARC Centre of Excellence in Future Low-Energy Electronics Technologies, Monash University, 3800 Victoria, Australia

[3]School of Physics and Astronomy, Monash University, 3800 Victoria, Australia

[4]Instituto Madrileño de Estudios Avanzados en Nanociencia (IMDEA-Nanociencia), 28049 Madrid, Spain

[5]Instituto Nicolás Cabrera, Universidad Autónoma de Madrid, 28049 Madrid, Spain

[6]Condensed Matter Physics Center (IFIMAC), Universidad Autónoma de Madrid, 28049 Madrid, Spain

[7] Institut de Ciència de Materials de Barcelona, ICMAB-CSIC, Campus UAB, 08193 Bellaterra, Spain

**Table of contents**

- **1: Models of the CDW-induced lattice distortions.**
- **2: Super-periodic patterns at room temperature.**
- **3: Line profiles across different reconstructions.**
- **4: STS measurements.**
- **5: Relative coverage with temperature**

## 1. Models of the CDW-induced lattice distortions

Many bidimensional materials undergo Peierls distortions to form charge density wave (CDW), a static modulation of the conduction electrons, which results in the opening of total or partial gaps at the Fermi level and is therefore energetically favored. CDWs are usually accompanied by a distortion of the lattice itself.

In **Figure S1** we report cartoon models of the lattice distortions associated with the main CDW patterns described in the main text.

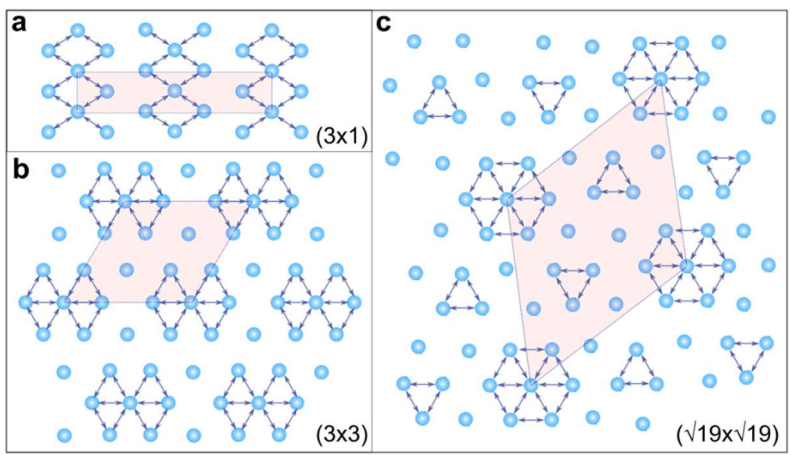

**Figure S1:** Charge density wave orders observed on our samples. Blue balls represent transition metal atoms in the TMD layer, blue arrows indicate the directions along which the lattice constant is reduced.

## 2. Evolution of the super-periodic patterns at room temperature

In this section, we report further STM data collected on the samples at room temperature.

**Figure S2** presents the bias-dependent modulation of the mixed periodicity pattern observed on TaTe$_2$ islands exhibiting the octahedral phase, described in Figure 2 in the main text. The images have been collected subsequently and on the same area, as demonstrated by the couple of defects present in every frame. While at low biases (closer to the surface, middle row) the coexistence of two different super-periodicities in the top and bottom part of each image is clear. We find that at large positive bias the modulation appears like an hcp lattice of bright spheres, while at large negative bias it appears like a mesh of wires arranged in a hexagonal pattern.

In **Figure S3**, we report the other super periodic patterns observed at room temperature on islands with the octahedral reconstruction. Figure S3a shows the flower-like reconstruction which has been attributed to a ($\sqrt{19}x\sqrt{19}$) CDW. Figure S3b shows a large-area scan on a sample with very high coverage (larger than a ML) obtained by extending the growth time up to 45 minutes. Areas enclosed within blue circles in panel b exhibit a ($3x1$) stripe-like reconstruction, identical to the one described in Figure 4d in the main text.

**Figure S4** presents the bias-dependent imaging of an area exhibiting a hexagonal reconstruction. The images have been acquired one after the other on the same area. Despite the noise associated with the measurements acquired at the lowest voltages, we observe that the bright/dark pattern does not change periodicity not appearance. We therefore attribute this modulation to a moiré pattern between TaTe$_2$ and the graphene layer underneath it.

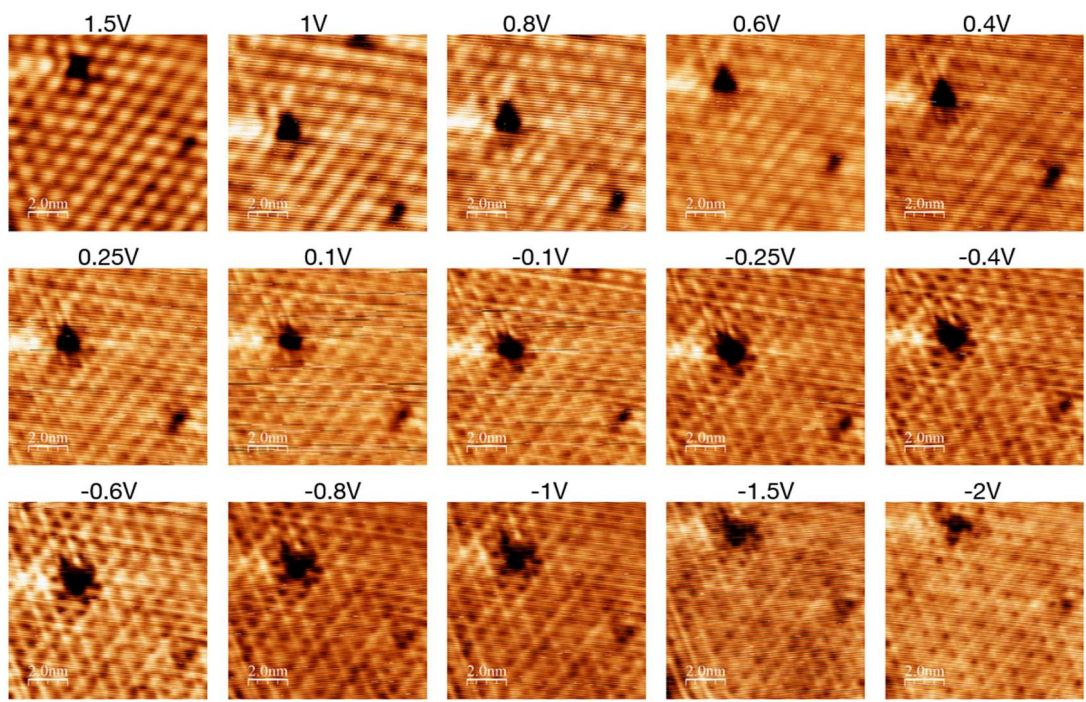

**Figure S2:** Evolution of the (octahedral) mixed (3x3)-(3√3/2 x4)R30° 1T CDW as a function of bias at RT. Scale bar = 2*nm*. I= 3*nA*.

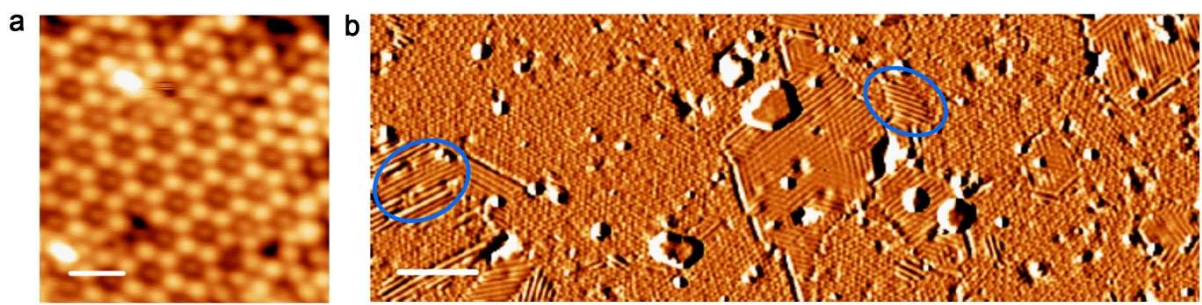

**Figure S3:** (a) Flower-like reconstruction, corresponding to a super-periodic pattern, observed at room temperature. V= -0.9*V*, I= 0.1 *nA*, Scalebar = 2 *nm*. (b) Derivative image of a large-scan area of sample with growth time 45 min, acquired at room temperature. Areas of the $(3x1)$ stripe-like reconstruction are highlighted by blue circles. V=1*V*, I= 0.1 *nA*, Scalebar=12 *nm*.

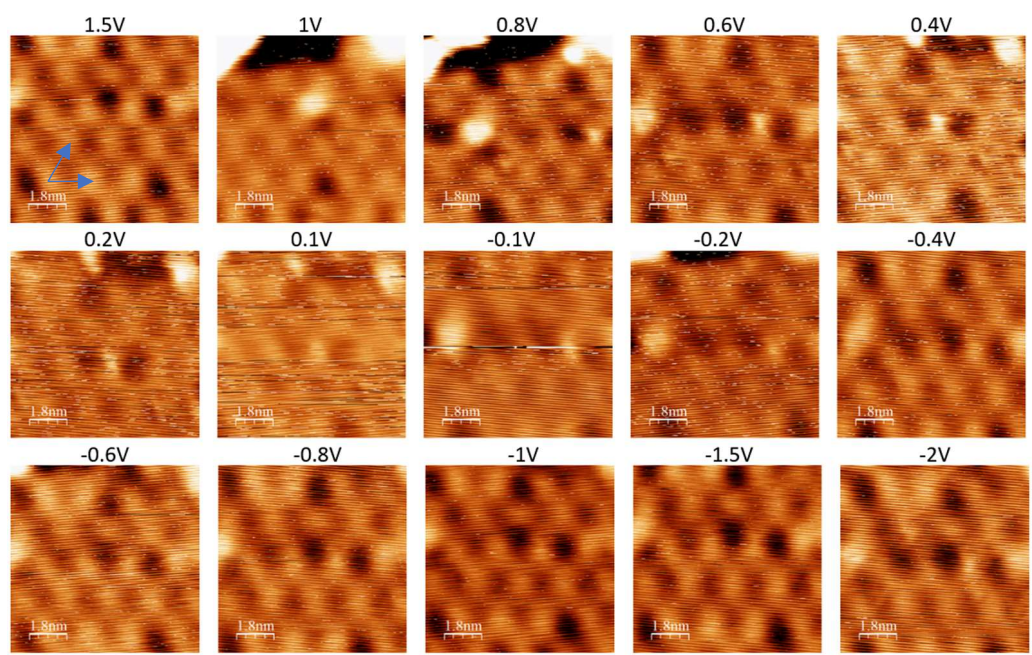

**Figure S4:** Evolution of the 1H moiré pattern of the 1H phase as a function of bias at RT. Scalebar: 1.8 nm. The moiré unit cell ($\sim 2.2\ nm$) is represented by blue arrows in the first panel. I= 3$nA$.

### 3: Line profiles across different reconstructions.

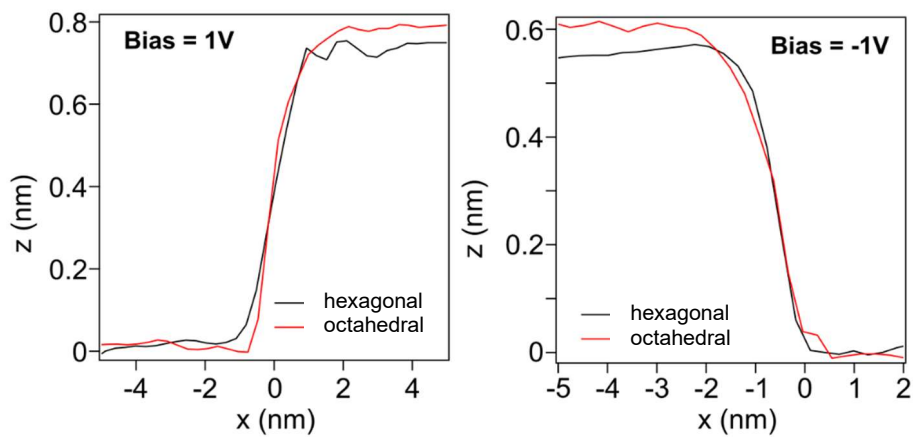

**Figure S5:** Exemplary line profiles acquired across the step edge of islands exhibiting a hexagonal (black lines) or octahedral (red lines) reconstruction. All measurements acquired at room temperature, I= 0.1 $nA$.

**4: STS measurements.**

The dip in DOS around Fermi level, reported in Figure 4e of the main text, is compatible with the existence of CDW pseudogaps and supports the attribution of these super-periodic modulations to CDWs. From the size of the pseudogap (~100 $meV$, obtained by performing a gaussian fitting across the dip and measuring the full width at half maximum) we can extract the CDW formation energy: knowing that $2\Delta = 3.52\ k_B T_{CDW}$, we obtain $T_{CDW} \approx 300\ K$. This is in line with the observation of CDW on the octahedral phase up to room temperature.

These observations point to the coexistence of multiple CDWs on the surface of the octahedral phase at the probed temperatures, rather than the conversion of one into another at given growth and annealing conditions, as reported by Hwang and coworkers.[21]

**5: Relative coverage with temperature.**

To explore the temperature dependence of the phase coverage, we performed several growths under the same experimental conditions except for the substrate temperature. The sample was sputtered clean after each growth and a pristine layer of graphene was grown on the Ir(111) substrate every time. We processed large-area STM images acquired for each growth (at least 9 images per sample, at least 150x150 $nm$ per image) and established the relative coverage of each phase with the Gwyddion software. The results are reported in **Figure S6**, and have been used to obtain the plot in Figure 6 of the main text.

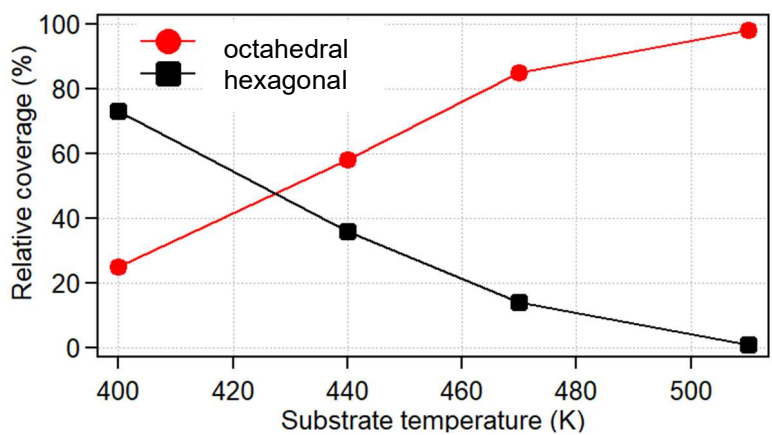

**Figure S6:** Relative coverage of the octahedral and hexagonal phases substrate temperature. All growths used for this graph have been performed with a 1:13 Ta:Te ratio and the growth time was set to 30 min.